\documentclass[12pt]{article}

\usepackage{graphicx,picins,a4}
\begin{document}
\title{``Measurement'' by neuronal tunneling: 
         Implications of Born's rule}
\author{L. Polley}
\date{\small Institut f\"ur Physik, Universit\"at Oldenburg, 26111 Oldenburg, FRG}
\maketitle
\begin{abstract}\noindent
A non-collapse scenario for ``conscious'' selection of a term from a 
superposition was proposed in quant-ph/0309166: thermally assisted 
tunneling of neuronal pore molecules. But ``observers'' consisting 
of only two neurons appear to be at odds with Born's rule.
In the present paper, an observer is assumed to possess a large number of 
auxilliary properties irrelevant for the result of the measurement. Born's rule 
then reduces to postulating that, prior to the result becoming conscious, 
irrelevant properties are in an entangled state with maximum likelihood, in the 
sense that phase-equivalent entanglements cover a maximal
fraction of the unit sphere (leading to equal-amplitude superpositions).   
\end{abstract}

\section{Introduction}

A persisting question of quantum theory is whether 
``measurement'' is a unitary process, essentially determined by a Schr\"odinger
equation with Coulomb interactions of electrons and nuclei, or whether completely
different physical or even non-physical structure is involved. 
Any unitary scenario of measurement will have to provide a mechanism for 
the stochasticity inherent in a quantum measurement, and will have to explain 
what happens to the discarded components of a wavefunction after collapse.

In \cite{Polley2003} a scenario was proposed in which a wavefunction
does not collapse but only part of it enters the consciousness of observers.
It was assumed that consciousness requires the firing of a neuron, and hence
\cite{Kandel2000} the transition of a channel-pore\footnote{In view of
the signal-amplifying role of chemical synapses in the brain, it is probably 
more appropriate to envision instead the opening of {\em fusion pore} molecules
(chapters 10 and 14 of \cite{Kandel2000}) which induce the release of 
neurotransmitters from synaptic vesicles. However, this does not affect the 
physical aspects of the scenario considered here. }  
molecule from a closed to an open state.
Such a transition as the basis of quantum measurement was extensively discussed
already by Donald \cite{Donald1990}. In \cite{Donald1990}, however, the switching
of pore molecules was assumed  to be an irreducible stochastic process, whereas 
in \cite{Polley2003} it was hypothesized that the transition involves molecular 
tunneling, assisted by a thermal environment. In this latter scenario, the 
microstates of neuronal heat baths, of all observers involved, {\em determine} 
the collective perception of the ``measured'' result.        

Some potential problems of the scenario were listed in \cite{Polley2003}.
One of them, which is the subject of the present paper, is apparently due to an  
oversimplified model of an ``observer''---assumed in \cite{Polley2003} to consist
of only two neurons plus heat baths. In such a model, the probability for 
obtaining the result $\mathrm{L}$ or $\mathrm{R}$ from a superposition 
$$
  a |\mathrm{L}\rangle + b |\mathrm{R}\rangle
$$ 
is not proportional to $|a|^2$ or $|b|^2$, respectively, but rather to something
like $|a|^{1/20}$ or $|b|^{1/20}$. This is in violation of Born's rule, except in
case of equal amplitudes $|a|=|b|$. The latter exception suggested an approach 
analogous to a derivation of Born's rule \cite{Deutsch1999,Zurek1998} from 
equal-amplitude superpositions of an extended system with many auxilliary states.
Obviously, in the nervous system of a realistic observer there are many 
candidates for auxilliary states. What remains to be provided for a 
complete picture of the measurement process is an argument as to {\em why\/} 
auxilliary states should systematically be superposed with equal amplitudes. 

The reason proposed here (section \ref{StatDom}) is the following.
A sufficiently complex observer, when engaged in an act of measurement, 
has a large number of nerval states available (ready to fire)
which are irrelevant for the result of the measurement but are nevertheless 
affected by the process. Unitary evolution, restricted to those available states, 
will be determined by some effective Hamiltonian. 
Because the states are ``irrelevant'', there is no reason for the measurement 
process to prefer any of them, so the process is likely to end up with a 
superposition belonging to a class with a large number of 
representatives.
Whether for a given class that number is large or small, relatively,
is uniquely determined since an invariant (basis-independent) measure on the unit
sphere is unique up to a proportionality 
factor. 
\footnote{An equivalent argument could presumably be formulated in terms of 
random matrices (evolution operators) drawn from the unitary circular ensemble 
\cite{Mehta1967}.}

An essential part of the scenario is that the response of an observer's nervous
system is not only determined by the object observed, but also by thermal
agitation of neuronal channel pore molecules. Thus, when the observer's system 
has interacted with the object, some neurons must be thermally induced 
to ``actually'' fire to produce a ``conscious'' result. This part of the 
measurement process has been explicated in \cite{Polley2003}. It was tested 
there numerically for an ``observer'' consisting of two neurons plus heat baths, 
so it remains to be generalized to an observer with many more neurons in 
ready-to-fire states. 
This is discussed in section \ref{Uniqueness}. 

Conclusions are given in section \ref{Concl}. An alternative, more intuitive 
way of counting representatives of classes of superpositions is given in the 
appendix.     

\section{Statistical dominance of equal amplitudes\label{StatDom}}

We wish to assign a statistical weight to certain classes of pure quantum states.
This requires a measure on the unit sphere in Hilbert space which is independent 
of the basis in which the states are represented. Such a measure is unique up to 
a multiplicative constant under very general conditions \cite{Sullivan1981}.

Let $n$ be the dimension of the Hilbert space, and consider a general state 
vector represented as
\begin{equation}  \label{GeneralSuperposition}
   |\psi\rangle = \sum_{k=1}^n c_k |k\rangle \qquad \qquad 
   \langle k | k' \rangle = \delta_{kk'} 
\end{equation}
The invariant measure is proportional to
$$
  \delta\left(1-\sum_{k=1}^n |c_k|^2\right) \prod_{k=1}^n \mathrm{d}^2 c_k
$$
Using polar coordinates so that $c_k = r_k e^{i\varphi_k}$, the measure takes 
the form
\begin{equation}  \label{InvariantMeasure}  
   \delta\left(1-\sum_{k=1}^n r_k^2 \right) \prod_{k=1}^n r_k \mathrm{d}r_k 
    \prod_{k=1}^n \mathrm{d}\varphi_k 
\end{equation}

\subsection{Volume of phase-equivalent superpositions}
   
Let us define
$|\psi'\rangle  = \sum_{k=1}^n c'_k |k\rangle $ to be {\em phase-equivalent} to 
$|\psi\rangle$ if there exist phase rotations of its coefficients such that 
$\| |\psi'\rangle_\mathrm{rotated} - |\psi\rangle \| < \epsilon$.
Calculationally, this reduces to a condition on the 
absolute values of the  coefficients:
\begin{equation}  \label{AmplitudeCondition} 
  \sum_{k=1}^n \left( r'_k - r_k \right)^2  < \epsilon^2 
\end{equation}
The reason for not requiring exact coincidence of absolute values here 
is that phase-equivalent sets would then all be of measure zero, so their sizes
could not be compared. An alternative argument, based on discretization instead 
of a normalization margin, is given in appendix \ref{IntuitiveNumber}.  

The calculation below will simplify if we, consistently, replace the delta 
function of equation (\ref{InvariantMeasure}) by the step function 
$(4\epsilon)^{-1} \chi\left(1-2\epsilon\leq\sum r _k^2\leq 1+2\epsilon\right)$.
The ensuing constraint is automatically satisfied for the primed radii 
if $\langle\psi|\psi\rangle = 1$ (because of triangle inequalities).

We now calculate $V_\mathrm{equiv}$, the volume on the unit sphere covered by  
superpositions phase-equivalent to of $|\psi\rangle$. Using the integral 
measure (\ref{InvariantMeasure}), we first obtain a factor of $(2\pi)^n$ by 
integrating over the phase angles. 
In the radial integrations, we substitute $r_k'$ by $s_k = r'_k-r_k$ and,
using $\epsilon \ll 1$, replace the factors $r'_k$ by $r_k$. The remaining 
integral equals the volume of an $n$-dimensional ball of radius $\epsilon$. 
Thus we obtain
\begin{equation} \label{Vphe}
   V_\mathrm{equiv} = 
   \frac{1}{4\epsilon} (2\pi)^n 
   \left( \prod_{k=1}^n r_k \right) V_\mathrm{ball}(n,\epsilon) 
  = \mathrm{const} \times  \prod_{k=1}^n |c_k|
\end{equation}

\subsection{Maximum likelihood for equal amplitudes\label{ML}}

The maximum of $V_\mathrm{equiv}$ given in (\ref{Vphe}), under the constraint
of unit normalization, is determined by the equation
$$
    \prod_{k=1}^n |c_k| - \lambda \sum_{k=1}^n |c_k|^2 = \mathrm{extremum}
$$  
Since all $c_k$ enter the same way, the obvious result is
$$
   |c_k| = \sqrt{\frac1n} \quad\mbox{for all }k
$$ 

\subsection{Fluctuations\label{MLfluc}}

Let us anticipate that root-mean-square deviations from a uniform amplitude 
configuration are small so that we need to take into account only first and 
second orders. Define relative deviations $\delta_k$ by
\begin{equation} \label{sqrt1/n+delta}
    |c_k| = \sqrt{\frac1n} \Big(1 + \delta_k\Big)  \qquad\qquad 
    {\cal O}(\delta_k^3) \mbox{ negligible}    
\end{equation}
The constraint of unit normalization implies
$$
   \sum_{k=1}^n  \delta_k = - \frac12 \sum_{k=1}^n \delta_k^2
$$ 
Hence, taking the logarithm of the number of phase-equivalent superpositions
and expanding in $\delta_k$, we obtain
$$
   \sum_{k=1}^n \log \Big(1 + \delta_k\Big)  = 
   \sum_{k=1}^n \delta_k - \sum_{k=1}^n \delta_k^2 + {\cal O}(\delta_k^3)
  =   - \frac32 \sum_{k=1}^n \delta_k^2
$$ 
Thus, the probability of relative {\sc rms} deviation 
$\delta =\sqrt{\frac1n\sum_{k=1}^n \delta_k^2}$
is a Gaussian of width $(3n)^{-1/2}$, 
\begin{equation} \label{probRMS}
  p(\delta) \propto \exp\left( - \frac32 \, n \, \delta^2 \right)
\end{equation} 

\subsection{Further constraints: Conservation of observables}

Let us assume that the observer's nervous system, when engaged in an act of 
measurement, has $n$ basis states available for reaction. 
Unitary evolution, restricted to those states, will be determined by some 
effective Hamiltonian.   
Let us assume the properties of those states to be ``irrelevant'' in 
the sense that the result of the measurement is solely contained in a tensorial 
factor $|\mathrm{L}\rangle$ or $|\mathrm{R}\rangle$ associated with each of the 
$n$ nerval states. Thus, prior to thermal agitation of the channel pore molecules, 
the entangled state is of the form
\begin{equation} \label{SuperpositionLR}
   \sum_{k=1  }^m c_k |k\rangle |\mathrm{L}\rangle 
 + \sum_{k=m+1}^n c_k |k\rangle |\mathrm{R}\rangle 
\end{equation} 
where nerval states associated with $|\mathrm{L}\rangle$ have been relabeled so 
as to appear as the first $m$ summands. 

Measurement should not change the value of the measured quantity. Hence, let us 
assume that the unitary operator of the process commutes with the observable. 
If the original state of the quantum system is 
\begin{equation} \label{}
    a |\mathrm{L}\rangle + b |\mathrm{R}\rangle
\end{equation} 
we must impose on the coefficients of (\ref{SuperpositionLR}) the constraints
\begin{equation} \label{NormConstraint}  
  \sum_{k=1  }^m |c_k|^2 = |a|^2 \qquad\qquad
  \sum_{k=m+1}^n |c_k|^2 = |b|^2
\end{equation} 
We already know from sections \ref{ML} and \ref{MLfluc} that amplitude 
configurations with maximum likelihood under the constaints 
(\ref{NormConstraint}) are characterized by
\begin{equation} \label{cLcR}
  |c_k| = \frac{|a|}{\sqrt{m  }} \quad k=  1,\ldots,m  \qquad\qquad
  |c_k| = \frac{|b|}{\sqrt{n-m}} \quad k=m+1,\ldots,n
\end{equation}  
Thus, by equation (\ref{Vphe}), the number of phase-equivalent configurations
is proportional to
$$
   \frac{|a|^m}{(\sqrt{m})^m} \, 
   \frac{|b|^{n-m}}{(\sqrt{n-m})^{n-m}}
$$ 
The logarithmic derivative of this expression with respect to $m$ ist
$$
    - \frac12 \log m + \frac12 \log (n-m) + \log |a| - \log |b|
$$
which is zero for
$$
    \frac{|a|}{\sqrt{m}} = \frac{|b|}{\sqrt{n-m}} 
$$ 
Hence, by (\ref{cLcR}), all $|c_k|$ are equal. We have thus shown that the 
equal-amplitude superpositions considered by Deutsch \cite{Deutsch1999} and 
Zurek \cite{Zurek1998} are distinguished by their statistical dominance in a 
sufficiently complex unitary scenario. 

Deviations from equal amplitudes have the probability determined in 
section \ref{MLfluc}. The probability tends to zero with $n\to\infty$. 

\section{Selectivity\label{Uniqueness}}

The analytical and numerical study of \cite{Polley2003} showed that thermally 
assisted neural tunneling (modeled with a hypothetical choice of parameters) 
is capable of selecting a single term from a superposition of two terms---the 
firing amplitude of one neuron highly exceeds the firing amplitude of the other 
neuron in the majority of cases. 

In the present, more complex scenario a single term would have to be selected 
from a superposition of $n$ quantum states of the observer's nervous system. 
Peculiar to extreme-value statistics (Gumbel distribution), this does not
imply a dramatic decline in selectivity even with $n\to\infty$.

Let us recall that in \cite{Polley2003} the distribution function for extremal 
values $\Phi$ of elongations of the pore molecule was approximated by 
the Gumbel form \cite{Embrechts1997}     
\begin{equation} \label{F(Phi)}
 F(\Phi) 
     =\exp\left(-\exp\left(-\frac{\Phi-\mu}{\sigma}\right)\right)
\end{equation}
with $\sigma$ parameterizing the thermal fluctuations induced by the neuronal 
heat bath\footnote{To the extent that local elongations $\phi(\vec{s},t)$
of a harmonically oscillating phononic heat bath can be regarded as statistically 
independent at sampling times $t_1,\ldots,t_N$, the Gumbel form is the exact 
limit distribution of the maxima for $N\to\infty$. We are assuming 
$n\ll N$ here.}.  
The maximum of the tunnel amplitude was proportional to $\exp(\kappa\Phi)$ with 
$\kappa$ deriving from the parameters of the internal molecular tunnel barrier. 

In the approximation (\ref{F(Phi)}) we can easily obtain the probability for 
the largest elongation in an ensemble of $n$ draws of $\Phi$ to differ from 
the second-largest elongation by an amount greater than $a$ 
(some chosen margin of selectivity). We have to
sum the probabilities for $\Phi_k$ ($k=1,\ldots,n$) taking some value while all 
other elongations are smaller than $\Phi_k - a$. This gives
\begin{equation} \label{P(n,a)}
  p(n,a) = n \int \mathrm{d}F(\Phi) \, \Big( F(\Phi-a) \Big)^{n-1}
  = \frac1{\left(1-\frac1n\right)e^{a/\sigma} + \frac1n}
\end{equation}
using $F(\Phi-a) = \big(F(\Phi)\big)^{e^{a/\sigma}}$ and integrating over the 
variable $F$ from $0$ to $1$. In particular, we recover the result 
$p(2,a)=1-\tanh(a/2\sigma)$ of section 3.2 of \cite{Polley2003}.  

Aiming at the selection by an outstanding tunnel amplitude of a term in a 
superposition, we are interested in cases where the probability (\ref{P(n,a)}) 
is close to unity, so that $a/\sigma\ll 1$. To first order in $a/\sigma$ we have
$$
    p(n,a) \approx 1 - \left(1-\frac1n\right)\frac{a}{\sigma}
$$ 
Hence, for arbitrary $n$ the probability for insufficient selectivity,
$1 - p(n,a)$, cannot be more than twice that fault probability with $n=2$
(which case was illustrated in Figure 3 of \cite{Polley2003}).

\section{Conclusion\label{Concl}}

In continuation of \cite{Polley2003}, the present approach aims to recover the 
dynamical pattern of quantum measurement in the ``conscious'' parts of some
unitary, Schr\"odinger-type dynamics of a suitably constituted ``observer''.
One of the potential problems listed in \cite{Polley2003} has 
been eliminated---Born's rule is satisfied if an ``observer'' consists of many 
more than two neurons. The solution of the problem, utilizing the large number 
of ``auxilliary'' states of such a nervous system,
was suggested by an elementary derivation of Born's rule in \cite{Deutsch1999}
and \cite{Zurek1998}. Two points were to be demonstrated in the present paper:
\begin{itemize} 
\item The {\em most likely superpositions\/} of 
auxilliary states have equal amplitudes.
\item Thermally assisted tunneling of neuronal pore molecules leads to 
{\em dominance of one summand} in a superposition of many neuronal states with 
the same efficiency as with only two states.        
\end{itemize}
This was shown on the basis of fairly standard results:
Uniqueness of the unitarily invariant measure on the unit sphere, and the Gumbel 
distribution for extreme values of Gaussian fluctuations.   

The unitary structure of Hilbert space was taken for granted here. 
In particular, state vectors were assumed to be normalized in the usual
way. It should be noted that this does not imply a problem of circularity
in the present context. We did not undertake to derive Born's rule but to show
that it is consistent with {\em unitary\/} quantum-mechanical time evolution of 
a system which has some relevant physical structure in common with a conscious 
observer engaged in a measurement.

\appendix

\section{Statistics of discretized amplitudes\label{IntuitiveNumber}}

\parpic(55mm,80mm)(0mm,55mm)[r]
{\includegraphics[width=55mm,keepaspectratio]{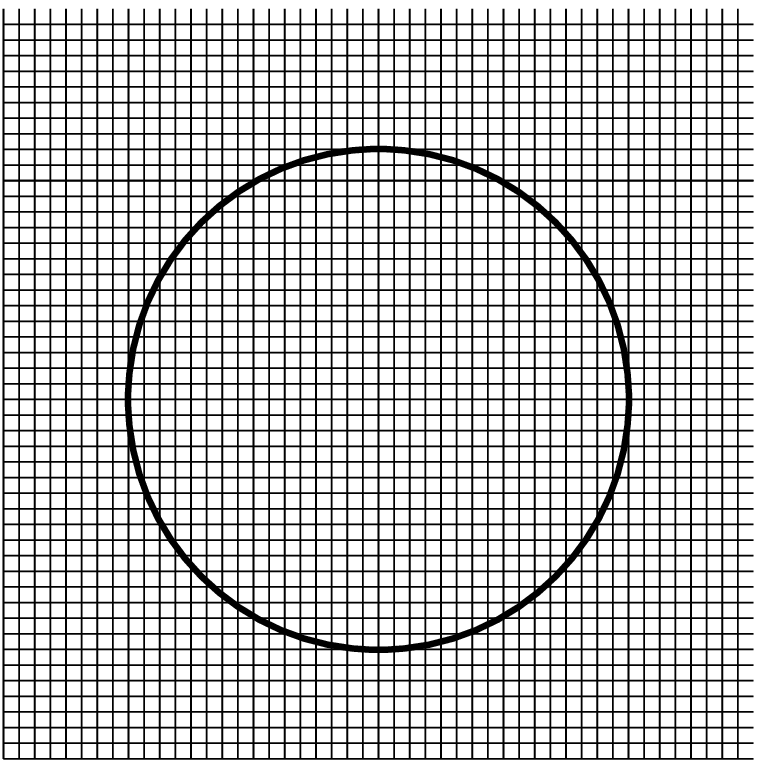}}
\noindent Let us divide the plane of complex probability amplitudes into squares
of length $d$ in the real and imaginary direction. 
The number of squares traced by a circle of radius $r_k$ is, 
in average over a small range of radii, $2\pi r_k/d$. Hence the total number 
of $2n$-dimensional cells representing phase-equivalent superpositions,
$$
    N_\mathrm{equiv} = \left[ \frac{2\pi}{d}\right]^n \prod_{k=1}^n |c_k| 
$$ 
Equation (\ref{Vphe}) is thus recovered (with a very large proportionality 
factor). 


\end{document}